\documentclass[twocolumn,english,prl,showpacs,preprintnumbers,superscriptaddress]{revtex4-1}
\usepackage[utf8]{inputenc}
\usepackage{color}
\usepackage{amsmath}
\usepackage{amssymb}
\usepackage{graphicx}
\usepackage{textcomp}
\usepackage{dcolumn}
\usepackage{bm}
\usepackage[right]{eurosym}
\usepackage{float}
\usepackage[english]{babel}
\usepackage{blindtext}
\usepackage{babel}

\begin{document}

\title{A simple photoionization scheme for characterizing electron and ion spectrometers}

\author{A. Wituschek}
\affiliation{Physikalisches Institut, Universit{\"a}t Freiburg, 79104 Freiburg, Germany}
\author{J. von Vangerow}
\affiliation{Physikalisches Institut, Universit{\"a}t Freiburg, 79104 Freiburg, Germany}
\author{J. Grzesiak}
\affiliation{Physikalisches Institut, Universit{\"a}t Freiburg, 79104 Freiburg, Germany}
\author{F. Stienkemeier}
\affiliation{Physikalisches Institut, Universit{\"a}t Freiburg, 79104 Freiburg, Germany}
\author{M. Mudrich}
\email{mudrich@physik.uni-freiburg.de}
\affiliation{Physikalisches Institut, Universit{\"a}t Freiburg, 79104 Freiburg, Germany}

\begin{abstract}
We present a simple diode laser-based photoionization scheme for generating electrons and ions with well-defined spatial and energetic ($\lesssim 2$ eV) structures. This scheme can easily be implemented into ion or electron imaging spectrometers for the purpose of off-line characterization and calibration. The low laser power $\sim 1$ mW needed from a passively stabilized diode laser and the low flux of potassium atoms in an effusive beam
make our scheme a versatile source of ions and electrons for applications in research and education.
\end{abstract}

\date{\today}

\maketitle
\section{Introduction}
Imaging detection of electrons and ions is an established technique in modern experiments dedicated to photoionization and chemical dynamics studies of gas-phase systems using pulsed lasers, synchrotron or free-electron laser radiation~\cite{Garcia:2009,Bozek:2009,Strueder:2010,Keeffe:2011,Lyamayev:2013}. Applications of imaging techniques range from single atoms~\cite{Skruszewicz:2014} and molecules~\cite{Suzuki:1999}, over clusters~\cite{Bartels:2009}, to nanoparticles~\cite{Zherebtsov:2011}. The advantage of electron and/or ion imaging over other detection techniques is the great level of detail with which light-matter interaction processes can be revealed, including ion fragment masses and velocities, photoelectron energies, as well as ion and electron angular distributions. By detecting electrons and ions in coincidence, ion mass-resolved photoelectron spectra and angular anisotropies or even the full kinematics of the ionization process can be recorded~\cite{Garcia:2009,Keeffe:2011,Ullrich:2003}.

However, photoionization experiments often rely on broad-band femtosecond lasers with a limited tuning range of the center wavelength, which may not be well adapted to performing precise electron and ion energy calibration measurements. In photoionization experiments at facilities such as synchrotrons or free-electron lasers usually the radiation is not permanently available as it is shared between several end-stations and working groups. 
This makes it necessary to develop procedures for testing and calibrating the detector systems as well as possible off-line. Besides, these sources of radiation are often limited in their tuning range towards low photon energies. In that case it may be beneficial to complement detector characterizations with laser-based measurements. However, the usage of intense lasers in such open environments, if available, is often hampered by strict safety regulations. 

Therefore, in this paper we present a simple yet very efficient photoionization scheme based on a low-power diode laser in combination with a dilute effusive beam of potassium (K) atoms. Moreover,
we suggest a rather simple setup that can be used for demonstration purposes and as students' laboratory project when combined with our photoionization scheme. With this scheme, both electrons and ions are generated with well-defined energies and angular distributions in a range of energies below 2 eV. Note that the maximum kinetic energy accepted by a typical velocity-map imaging (VMI) spectrometer scales linearly with the electrode voltages~\cite{Eppink:1997}. Thus, to a certain extent a spectrometer designed for a different energy range can be characterized with the presented scheme as well. 

\begin{figure}[hbt]
\center
\includegraphics[width=0.6\columnwidth]{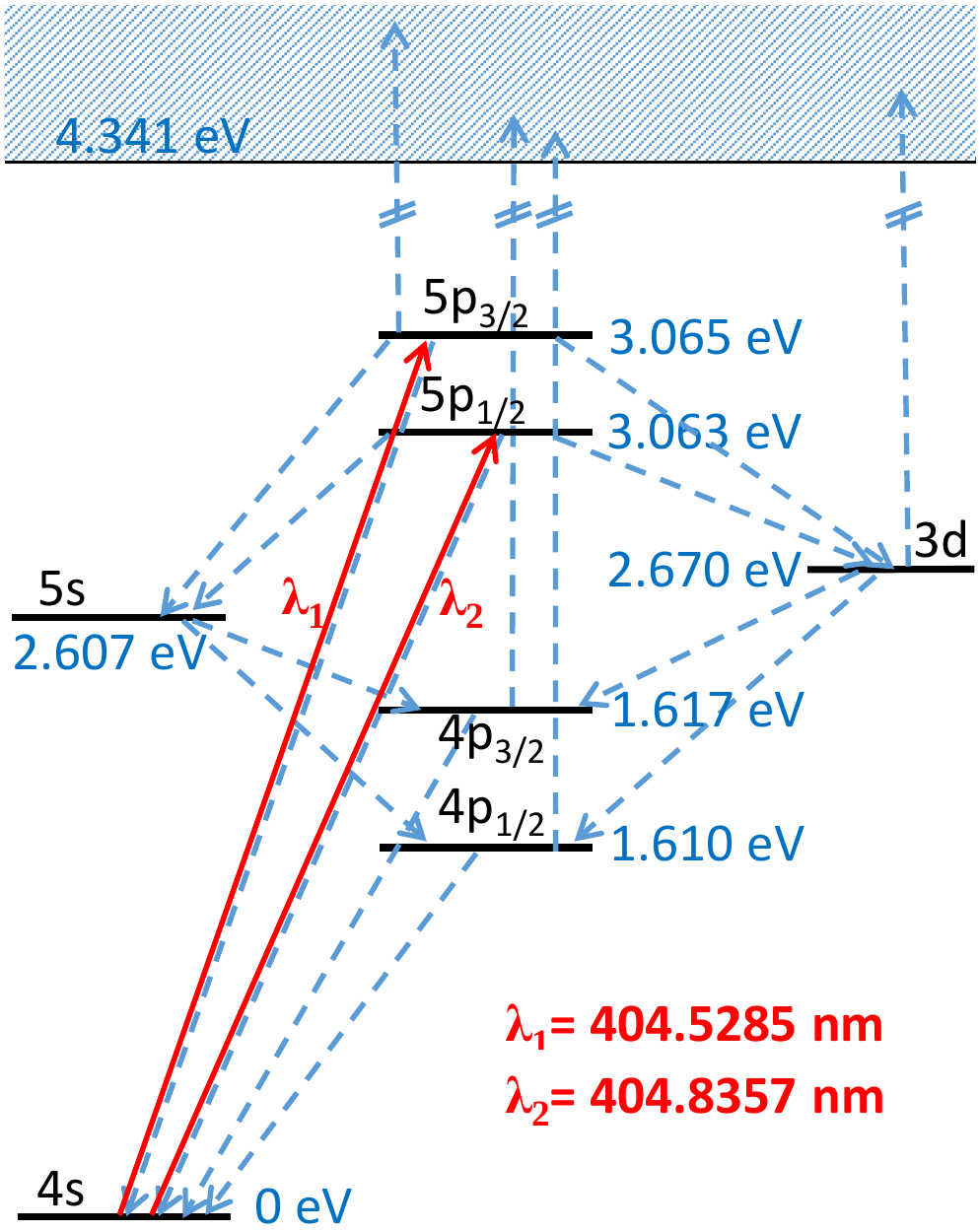}\caption{\label{fig:levels} Level diagram of potassium including the relevant atomic levels. The dashed lines represent the relevant allowed optical transitions.}
\end{figure}
\section{Experimental implementation}
The choice of the ionization scheme presented in Fig.~\ref{fig:levels} is based on the following requirements. Most importantly, a resonant low-order multiphoton process reaching up into ionization continuum with high transition cross sections for excitation and ionization is beneficial for achieving high ionization rates at low laser power. Ideally, the final ionization step generates a series of sharp photoelectron kinetic energies by ionizing out of a number of excited states with starkly differing binding energy. The light needed for driving the transitions is supposed to be narrow-band and supplied by a readily available laser source. 

All these requirements are fulfilled for the chosen two-photon transition in K where the excitation takes place on the 4s$\rightarrow$5p transition and ionization occurs out of the levels 4p, 3d, and 5p. The laser light at the resonance wavelengths (vacuum) $\lambda_0=404.8357$~nm (D1-line) or $\lambda_0=404.5285$~nm (D2-line) is provided by low-cost, high-power laser diodes developed for applications of the Blu-ray Disc technology. Narrow-band, frequency-controllable operation of diode lasers can be achieved using standard setups for instance using external grating stabilization~\cite{Ricci:1995,Cook:2012}. 

Besides the suitable level scheme and the availability of low-cost lasers, K is advantageous with regard to handling and creating an atomic beam. Due to the high vapor pressure at low temperatures, an effusive beam with a density of the order of $10^8$ atoms per cm$^{-3}$ in the ionization region can be generated from a sample of elementary K metal using a vapor cell heated to a temperature around $T_\mathrm{K}=140^\circ$C~\cite{Alcock:1984}. Alternatively, K vapor can be created using commercially available alkali metal dispensers (e.\,g. Saes Getters S.p.A.).

\begin{figure}[hbt]
\center
\includegraphics[width=1.0\columnwidth]{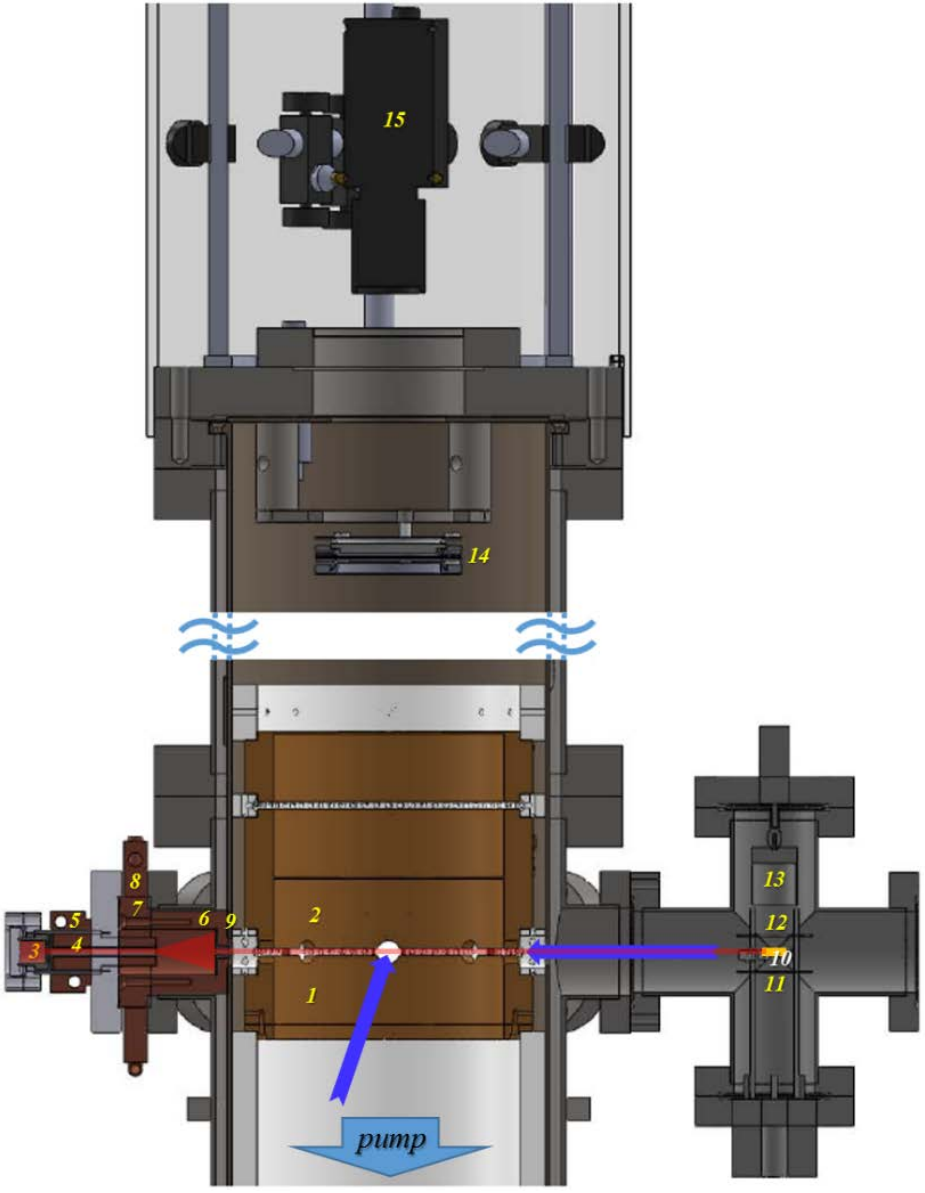}\caption{\label{fig:setup} Schematic representation of the experimental setup used in this work. The blue arrows indicate the possible paths of the laser beam.}
\end{figure}
For demonstrating the suggested ionization scheme we have set up a VMI spectrometer as shown in Fig.~\ref{fig:setup}. One side flange of the main chamber on the level of the laser ionization region, which is located between repeller (1) and extractor (2) electrodes, contains a heated K vapor cell for creating an effusive K beam as indicated by the red shaded area. A simple and compact design is chosen which gets along with no additional pumping nor electrical or other vacuum feedthroughs. Briefly, the K effusive beam source is incorporated into a standard 16~mm to 40~mm conflat flange (CF) adapter. Thus, with our design a well-collimated effusive K-beam is easily implemented into any vacuum chamber that provides a 40~mm CF or larger flange from which a free line of sight connects to the ionization region. In particular the connecting flange to the main source of radiation could be used when working off-line.

A small K reservoir is realized by welding a steel cup with a central 3~mm bore hole onto a 16~mm blank CF (3). An adjacent copper cylinder with an inner 3~mm steel channel of length 60~mm (4) is pressed into the tube of the flange adapter. Heating of the channel and reservoir is achieved by attaching a heated copper clamp from outside onto the flange adapter tube (5). A cold volume behind the heated channel is realized by bolting a copper cup with a central 2~mm bore hole (6) onto a copper ring (7). The latter serves as a heat-conducting support of the cold copper cup and at the same time as a gasket that connects the 40~mm CF to the main chamber. Cooling of that part is provided by a water or air cooled copper clamp attached to this copper ring from outside the vacuum (8). In this way only those K atoms emitted out of the heated channel that travel along the beam axis are transmitted through the hole in the cold copper cup to form an effusive beam, whereas the divergent part of the K flux out of the channel is mostly frozen out on the walls of the cold copper cup. An additional 2~mm aperture is mounted further downstream onto the $\mu$-metal shield (9) that surrounds the spectrometer electrodes. In this way, the effusive K beam is collimated to a diameter $< 10$~mm over the length of the spectrometer setup so that metallization of the surfaces of electrodes and isolators is safely prevented.

The four-way cross on the opposite side of the K vapor cell contains a simple surface ionization setup used for monitoring the K beam intensity~\cite{Delhuille:2002}. It consists of a resistively heated rhenium ribbon (10) placed between two oppositely poled plate electrodes (11, 12) and a steel cup which collects the K$^+$ charges generated from the K atoms impinging on the hot rhenium surface (13). The resulting ion current on the cup which is in the range of 5~nA is measured using a picoamperemeter.

The K beam is either intersected by the laser beam at right angles or counter-propagates with the laser beam coaxially inside the laser ionization region of the VMI spectrometer (blue arrows). The latter consists of a standard arrangement of three electrode plates and a grounded region of free flight, combined with a position-sensitive detector (MCP and phosphor screen) with an active area of diameter 40~mm (14)~\cite{Eppink:1997,Fechner:2012}. Either K$^+$ photoions or photoelectrons are accelerated by the electric field generated by the difference between voltages applied to the repeller ($U_{r}$) and the extractor ($U_e$) electrodes onto the detector area. The P43 phosphor screen is imaged through a glass window by a camera (Basler low-cost camera acA1920 containing a CMOS $1920\times 1200$ pixels monochrome sensor by Sony (1/1.2'' optical size) with a $f=25$~mm objective by Kowa (15). 

The laser used in this work is a commercial grating stabilized diode laser (DL 100 by Toptica) but other stable designs are equally suitable and readily available~\cite{Cook:2012}. The power of the laser beam at the entrance window of the spectrometer is $P_\mathrm{las}=3.5$~mW throughout this work. While the requirements on the laser line width are quite relaxed in this application ($<200$~MHz), good long-term (i.\,e. temperature) stability is needed in order to do without active frequency stabilization, as it is the case in this work. When setting the laser parameters (temperature, current, grating angle) for the first time so as to resonantly excite either of the two 5p levels of the K atom it is useful to measure the laser wavelength using a wavemeter capable of resolving the wavelength to a precision of $5\times 10^{-4}$~nm ($1$~GHz). 

Once the laser is preset in this way, focusing the laser beam in a counter-propagating fashion along the K beam will result in measurable ionization signals with only small variations of the laser parameters (current, grating angle). This is due to the large longitudinal Doppler broadening of the K absorption profile of $1.76$ GHz at a temperature of the K vapor cell of $T_\mathrm{K}=140^\circ$C. For the case of crossing the laser and K atomic beams at right angles the transverse Doppler broadening is limited to $\sim 220$ MHz by the divergence of the K beam. The latter is determined by the positions and diameters of the apertures along the K beam. Therefore, in implementations where the laser beam can only be made to intersect the K atomic beam near right angles the laser settings have to be slightly better predetermined. In that case, it is useful to monitor the resonance fluorescence emitted by K vapor in a separate heated ($130^\circ$C) vapor cell which can be commercially purchased (e.\,g. Toptica Photonics AG). 

In the crossed beams geometry with a laser beam radius $w=0.6$~mm and a focusing lens with focal length $f=150$~mm we measure a total ion count rate of about 30,000~counts/s by counting individual K$^+$ ion hits on the detector when tuning the laser onto the 5p$_{3/2}$ resonance at $\lambda=404.53$~nm and setting the K vapor cell to $T_\mathrm{K}=$140$^\circ$C. This high ionization rate considering the low laser power allows us to record ion and electron images with sufficient statistics within total exposure times of the camera of typically a few seconds.

\begin{figure}[hbt]
\center
\includegraphics[width=0.9\columnwidth]{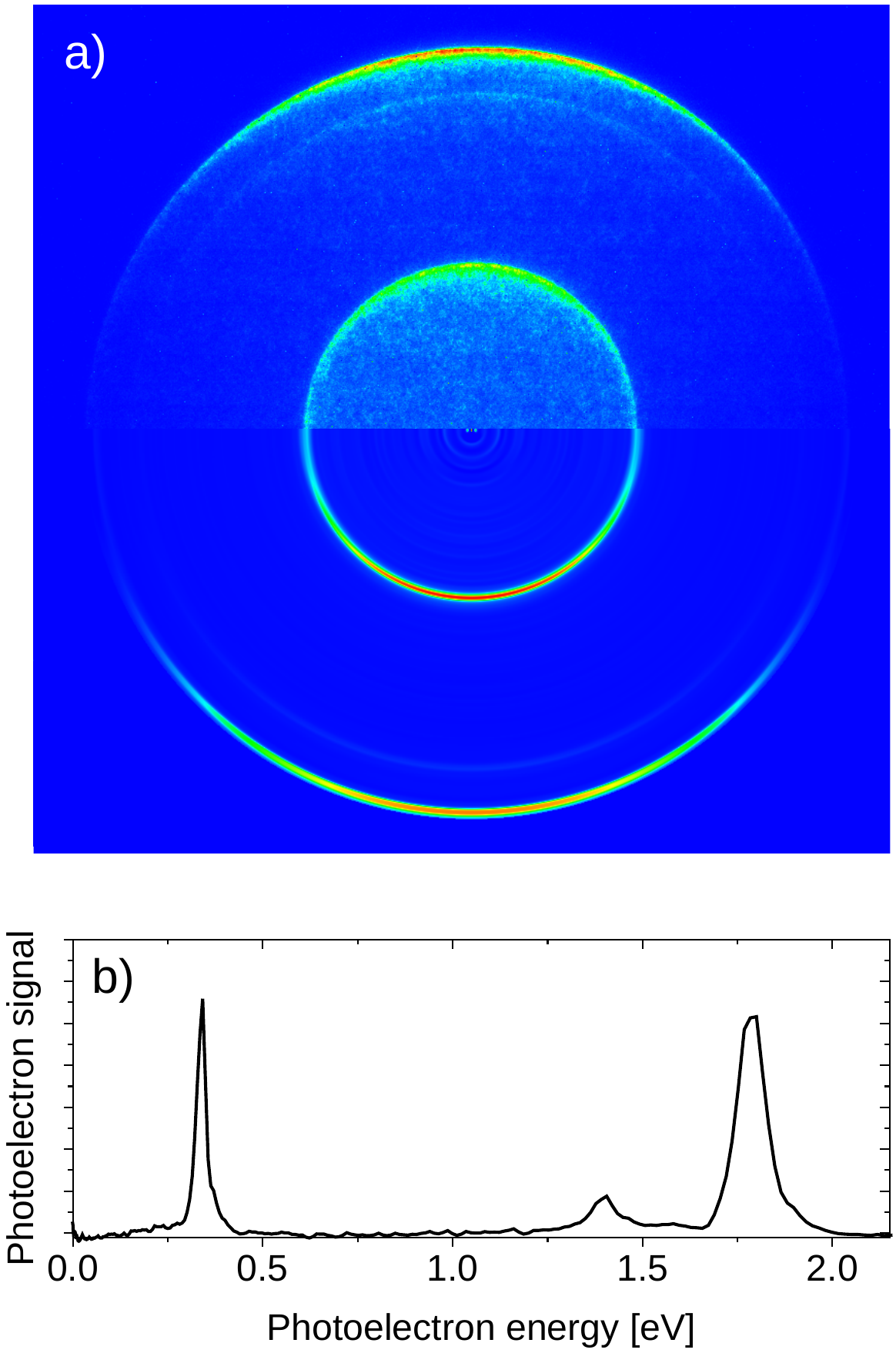}\caption{\label{fig:ElectronVMIblue} a) Raw photoelectron velocity-map image (upper half) and inverse Abel transformed image (lower half). b) Photoelectron spectrum inferred from the image.}
\end{figure}
\section{Electron imaging}
In the following we briefly present experimental results for different applications of the suggested photoionization scheme to characterizing our ion and electron imaging setup. We show that resolving powers as well as calibrations can be accurately determined for both spatial and velocity map imaging modes of operation. 

A typical VMI of photoelectrons is depicted in Fig.~\ref{fig:ElectronVMIblue} a). The upper half of the image shows the raw data, while the lower half is the inverse Abel transformed image obtained using the pBasex algorithm~\cite{Garcia:2004}. Note that for the full image reconstruction by inverse Abel transformation to apply the laser polarization must be aligned perpendicularly to the spectrometer axis. The experimental parameters are, $U_r=5$~kV and $U_e=3.6$~kV, $T_\mathrm{K}=150^\circ$C, and a total exposure time of the camera of 20~s. From this image we obtain the angle-integrated photoelectron spectrum (PES), shown in Fig.~\ref{fig:ElectronVMIblue} b), and the angular distribution for each feature in the PES which can be characterized by anisotropy parameters $\beta_2$ and $\beta_4$ for two-photon ionization processes~\cite{reid}. 

Clearly, the PES contains three sharp peaks which result from photoionization of the resonantly excited level 5p and of two lower-lying levels, 3d and 4p, which are populated by spontaneous decay out of 5p. The relative peak heights are determined by the spontaneous decay rates which can be found in the NIST atomic database~\cite{NIST} as well as by different photoionization cross sections of the individual levels~\cite{Moskvin:1963,Aymar:1976}. Photoionization of the 5s level is suppressed by a Cooper minimum in the photoionization cross section near the used photon energy~\cite{Moskvin:1963}. 

The peak positions correspond to the photoelectron kinetic energy values $E_e=h\nu +E_{n\ell} -E_i$, where $h\nu=hc/\lambda$ is the photon energy, $c$ is the speed of light, the level energies $E_{n\ell}$ and the ionization energy $E_i$ of K are given in Fig.~\ref{fig:levels}. These values can be used as calibration points for the photoelectron energy. The resulting calibration factor is $k_\mathrm{eVMI}=0.0133(5)$~meV/pix$^2$ for the given spectrometer settings. From the width of the peaks in the PES we determine the energy resolution of our spectrometer to $\Delta E_e/E_e =5$\,\%. Thus, the resolution is not sufficient for resolving the fine structure splitting of the 3d and 4p lines. For the anisotropy parameters we find the values $\beta_2^{5p3/2}=1.07(4)$, $\beta_4^{5p3/2}=0.52(8)$, $\beta_2^{3d}=0.86(9)$, and $\beta_2^{4p}=0.17(3)$. $\beta_4^{3d}$ and $\beta_4^{4p}$ are consistent with zero as expected for levels which are populated by spontaneous decay.


\begin{figure}[hbt]
\center
\includegraphics[width=0.9\columnwidth]{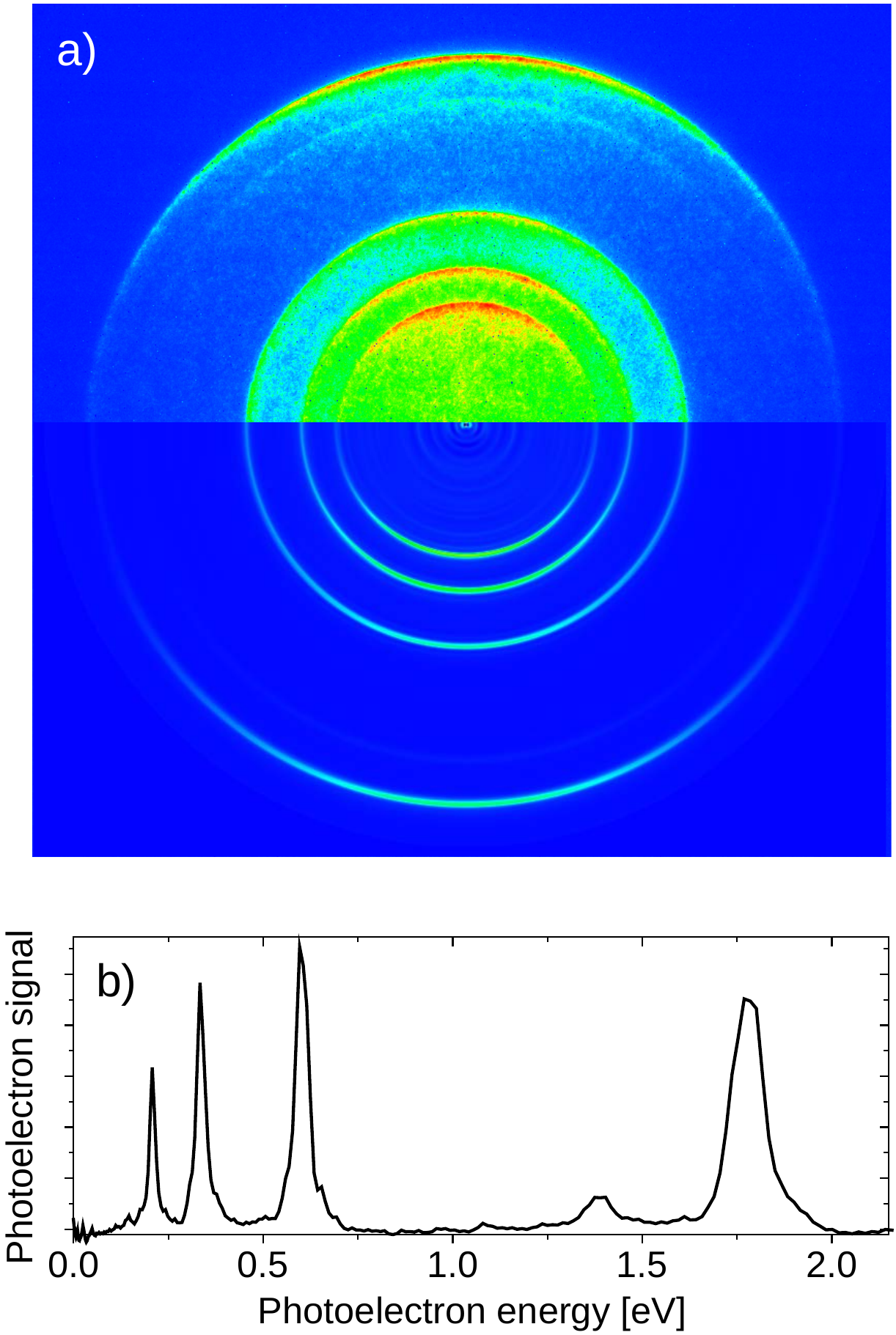}\caption{\label{fig:ElectronVMIred} a) Raw and inverted photoelectron image for photoionization using a blue and a red laser. b) Photoelectron spectrum inferred from the image.}
\end{figure}
We mention that the number of peaks in the PES can easily be increased by adding an additional laser with a different known wavelength, e.\,g. a helium-neon laser. Here we demonstrate the effect of superimposing an additional red diode laser ($\lambda_2=658.42$~nm) with a power of 12~mW coaxially with the blue laser in a counter-propagating fashion, see Fig.~\ref{fig:ElectronVMIred}. In addition to the peaks discussed previously, new features appear at $E_{e,2}=h\nu_2 +E_{n\ell} -E_i$. This increased number of peaks may be useful for obtaining a refined calibration curve.

\begin{figure}[hbt]
\center
\includegraphics[width=0.8\columnwidth]{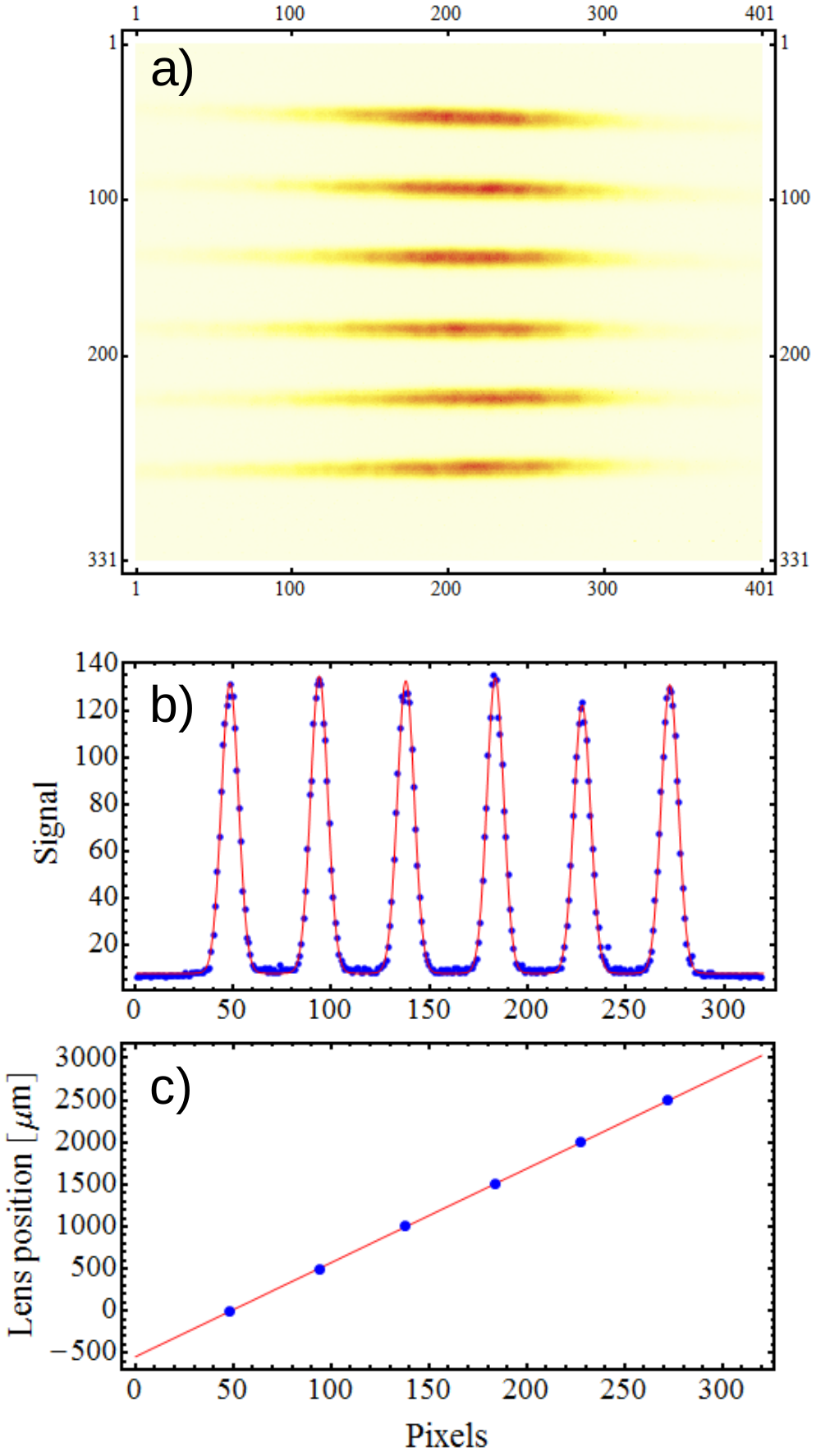}\caption{\label{fig:IonsSMI} a) Sum of spatial map ion images for positions of the laser focus displaced in steps of 0.5~mm. The laser beam intersects the K atomic beam at right angles. b) Fit of a vertical cut through a) by a sum of shifted gaussian functions. c) Linear regression of the peak positions.}
\end{figure}
\section{Ion imaging}
In the spatial map ion imaging (SMI) mode~\cite{Stei:2013}, for which we set the electrode voltages to $U_r=3$~kV and $U_e=2.7$~kV, the spatial resolution as well as the magnification scale can be determined by imaging K$^+$ ions generated by the tightly focused laser beam in the interaction region. The focus position is transversally displaced in steps of 500~$\mu$m by mounting the focusing lens ($f=150$~mm) onto a mechanical translation stage. The sum of the resulting ion images are shown in Fig.~\ref{fig:IonsSMI} a). In this mode of operation, care has to be taken not to saturate neither the imaging detector nor the camera because the signal is concentrated in a small area. 

A transverse cut through the two-dimensional intensity distribution along the centers of the elongated intensity maxima is displayed in Fig.~\ref{fig:IonsSMI} b) together with a least-squares fit of the data by the sum of shifted gaussian functions. From the resulting peak positions of the gaussians as a function of the lens position we determine the calibration factor $k_\mathrm{SMI}=11.2(2)~\mu$m/pix for the given voltage settings from a linear regression, see Fig.~\ref{fig:IonsSMI} c). The full width at half maximum (FWHM) of the gaussians, $\Delta x = 113~\mu$m, determines the resolution of our setup since the real FWHM of the transverse laser intensity distribution in the focus, $M^2 f\lambda_0/(\pi w)\sqrt{2\ln\left(2\right)}=41(2)~\mu$m, is much smaller than $\Delta x$~\cite{Eppink:1997,Stei:2013}. The measured limited resolution of our setup is attributed to aberrations of the ion optics.

\begin{figure}[hbt]
\center
\includegraphics[width=0.9\columnwidth]{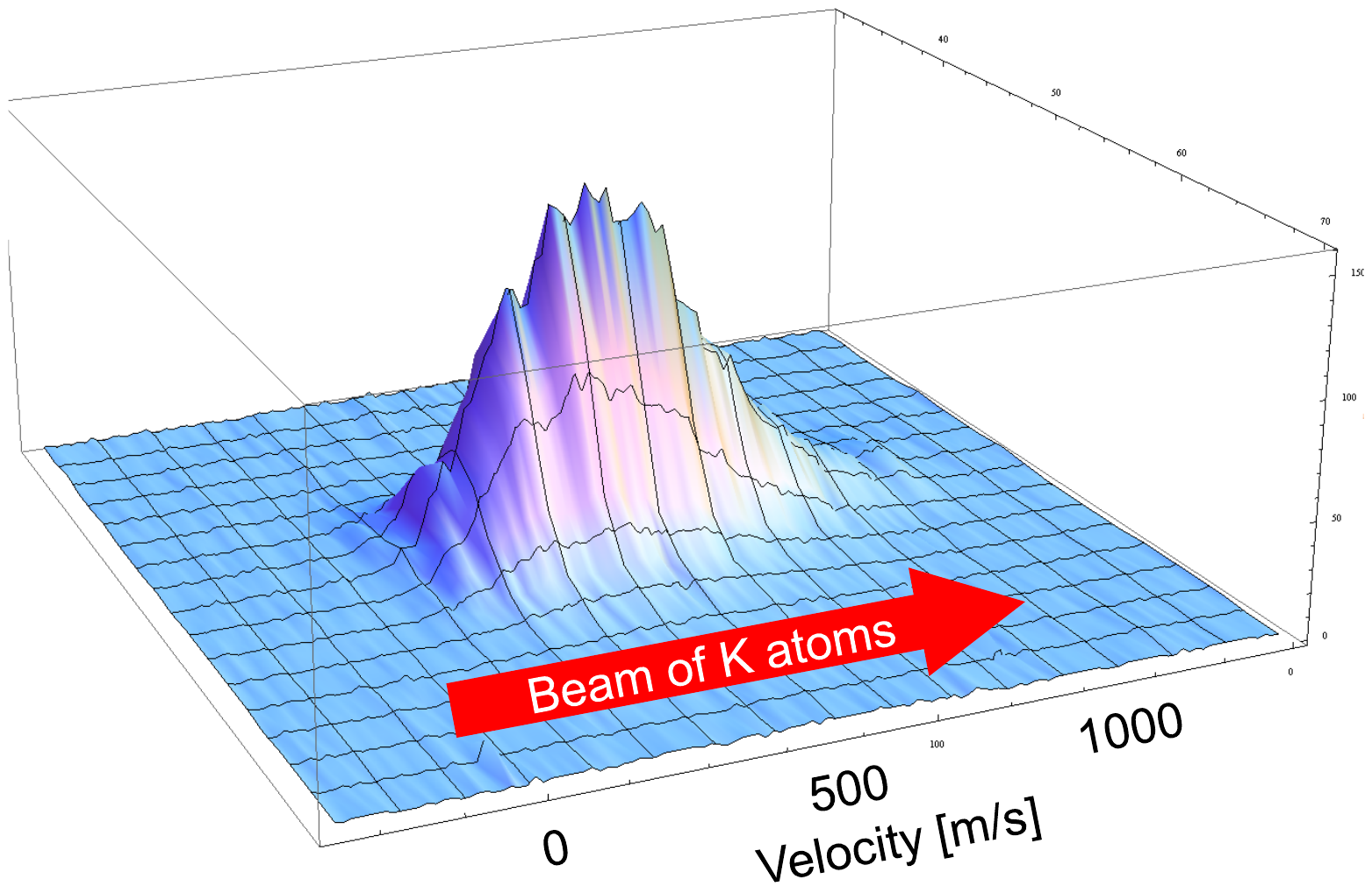}\caption{\label{fig:ionsVMItrans} Velocity distribution of K$^+$ ions imaged by crossing the laser and atomic beams at right angles.}
\end{figure}
In the ion VMI mode the full velocity distribution of the effusive K beam can be imaged when using the crossed beams geometry, see Fig.~\ref{fig:ionsVMItrans}. The coaxial beams geometry allows to calibrate ion velocities by imaging narrow velocity distributions which are individually addressed within the broad Doppler profile by precisely controlling the wavelength of our spectrally narrow laser ($\Delta\nu_\mathrm{las}\lesssim 1$~MHz). The laser wavelength is measured using a precise wavelength meter (High Finesse WS-7 MC4). Alternatively, a cheaper commercial or homebuilt Fabry-P{\'e}rot etalon can be used as an optical spectrum analyzer, since only the relative detuning of the laser is relevant for this measurement.

\begin{figure}[hbt]
\center
\includegraphics[width=1.0\columnwidth]{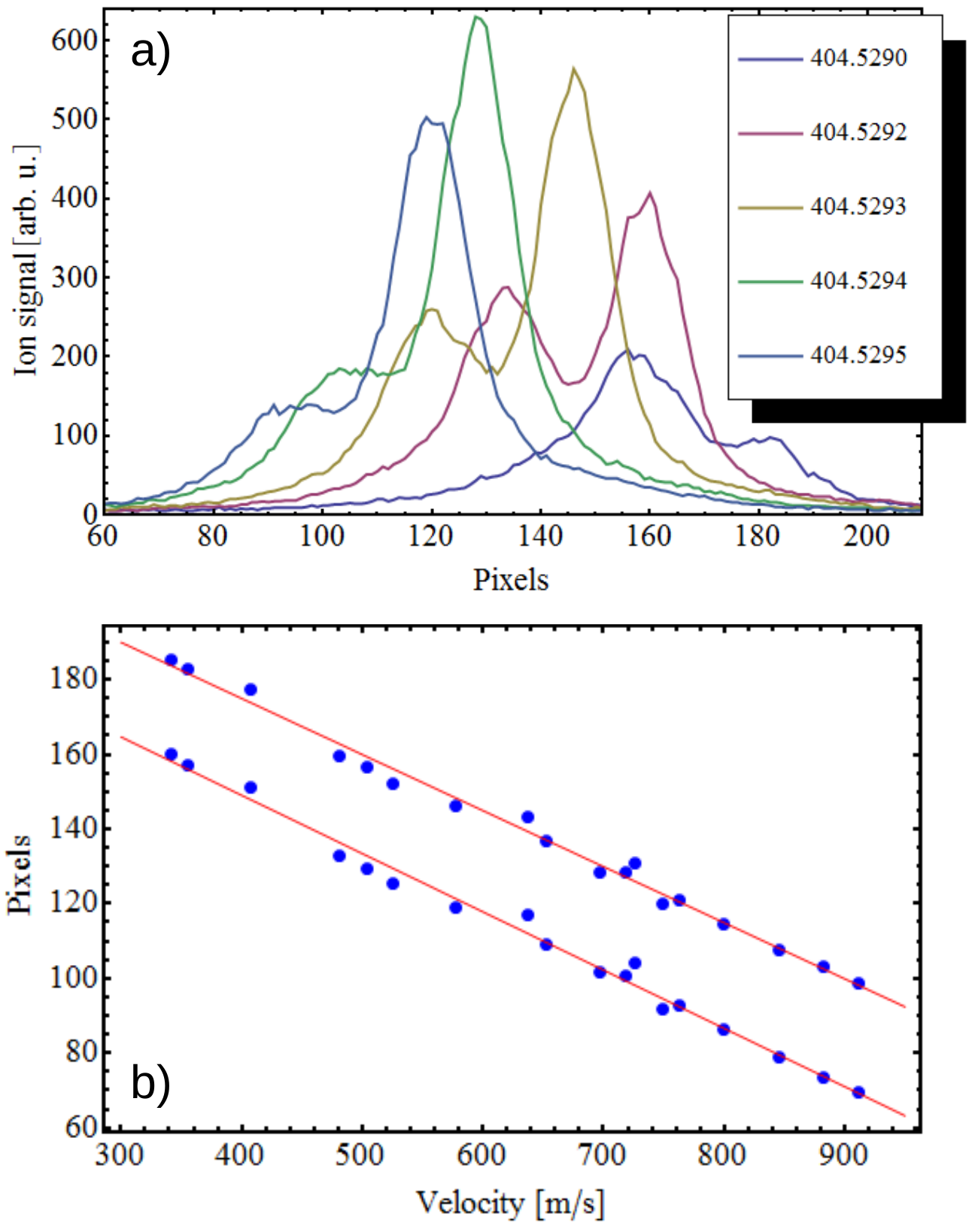}\caption{\label{fig:ionsVMI} a) Velocity profiles along the K atomic beam recorded at various laser wavelengths. The shifting is due to the Doppler effect, the double-peak structure reflects the hyperfine splitting of the K groundstate. b) Linear fits to the peak positions.}
\end{figure}
Fig.~\ref{fig:ionsVMI} a) shows cuts through the ion images along the direction of the K beam recorded at the indicated values of the laser wavelength.
The electrode voltages are set to $U_r=3$~kV and $U_e=2.2$~kV. The double-peak structure of the profiles directly reflects the hyperfine splitting of the groundstate of $^{39}$K (93\,\% natural abundance) which is $461.7$~MHz. These profiles are very well fitted by a sum of two shifted lorentzian functions. A linear regression of the detected peak positions as a function of ion velocity $v$ which is given by the laser wavelength $\lambda$ through the Doppler effect, $v=(\lambda_0/\lambda -1)c$ where $c$ is the speed of light, we obtain the calibration factor for K$^+$ velocities $k_\mathrm{VMI}=6.5(2)$~m/(s$\cdot$pix), see Fig.~\ref{fig:ionsVMI} b). We emphasize the benefit of this calibration procedure for ion velocities in the low energy range (15-150~meV) by a direct measurement which is hard to realize by other means. Photodissociation reactions which lead to well-known fragment velocities usually require intense ultraviolet lasers~\cite{Eppink:1997}. At higher energies in the eV range, the velocity calibration can be obtained using photoelectron imaging, as detailed in the preceding section. 

\section{Summary}
We demonstrate a simple yet efficient photoionization scheme that may be useful for characterizing and calibrating electron and ion spectrometers and imaging setups. The scheme is based on resonant two-photon ionization of potassium atoms in an effusive beam. A compact potassium atomic beam source and monitor are described. High ionization rates of $\sim 10$~kHz per mW of laser power are easily obtained using a passively frequency-stabilized diode laser. Due to the coincidence of the required laser wavelength ($~\sim 405$~nm) with that provided by the Blu-ray Disc technology, laser diodes with sufficient power are commercially available at low cost.

With this photoionization scheme, both electrons and ions are created with well-defined kinetic energies. Three or more electron energies in the range from 0.3 to 1.8~eV are generated by ionizing excited atoms using only the blue laser or additional lasers, respectively. Ions with controllable kinetic energies in the range of 15-150~meV can be generated by exploiting the Doppler-broadened absorption profile of atoms along the beam axis in combination with the narrow-band and well-controlled emission spectrum of the diode laser. Furthermore, the spatial mapping capabilities of ion imaging setups can be accurately characterized owing to the small and position-controlled ionization region given by the laser focus. Therefore we suggest this scheme to be implemented into setups for scientific, educational, or industrial applications, in particular where the main ionization process is not permanently available such as at synchrotrons or free-electron lasers. 

\acknowledgements{The authors gratefully acknowledge the generous donation of essential parts of the setup by Hanspeter Helm as well as support by the Deutsche Forschungsgemeinschaft in the frame of projects MU 2347/6-1 and IRTG 2079. We thank Fr{\'e}d{\'e}ric Vachon and Julian Baier for their help with the setup in the early stage of the construction.}


%

\end{document}